\documentclass{raa_twocolumn}            

\usepackage{graphicx,times}             
\usepackage{natbib}
\usepackage{amssymb,amsmath}
\bibpunct{(}{)}{;}{a}{}{,}

\usepackage[pagebackref=true]{hyperref}

\newcommand{\figref}[1]{Figure \ref{#1}}
\newcommand{\secref}[1]{Section \ref{#1}}

\newcommand{\exsitu}{\textrm {ex-situ}}
\newcommand{\insitu}{\textrm{in-situ}}

\begin{document}

  \title{Formation of super-thin galaxies in Illustris-TNG}

   \volnopage{Vol.0 (20xx) No.0, 000--000}      
   \setcounter{page}{1}          

   \author{Jianhong Hu \thanks{E-mail: hu.jianhong\_{}2008@163.com} 
   \and Dandan Xu \thanks{E-mail: dandanxu@tsinghua.edu.cn}
   \and Cheng Li
   }

   \institute{Department of Astronomy, Tsinghua University, Beijing 100084, China;
   \\
\vs\no
   {\small Received 20xx month day; accepted 20xx month day}}

\abstract{Superthin galaxies are observed to have stellar disks with extremely small minor-to-major axis ratios. In this work, we investigate the formation of superthin galaxies in the TNG100 simulation. We trace the merger history and investigate the evolution of galaxy properties of a selected sample of superthin galaxies and a control sample of galaxies that share the same joint probability distribution in the stellar-mass and color diagram. Through making comparisons between the two galaxy samples, we find that present-day superthin galaxies had similar morphologies as the control sample counterparts at higher redshifts, but have developed extended flat `superthin' morphologies since $z \sim 1$. During this latter evolution stage, superthin galaxies undergo overwhelmingly higher frequency of prograde mergers (with orbit-spin angle $\theta_{\rm orb} \leqslant 40^\circ$). Accordingly the spins of their dark matter halos have grown significantly and become noticeably higher than that of their normal disk counterparts. This further results in the buildup of their stellar disks at larger distances much beyond the regimes of normal disk galaxies. We also discuss the formation scenario of those superthin galaxies that live in larger dark matter halos as satellite galaxies therein.
}

   \authorrunning{J. Hu, D. Xu \& C. Li }            
   \titlerunning{Formation of super-thin galaxies in Illustris-TNG}  

   \maketitle

%
%

\section{Introduction}
\label{sec:intr}

Superthin galaxies were noticed by astronomers half a century ago for
their unusually thin and elongated shape \citep{voron1967, devau1974,
  goad1979, goad1981}. They are viewed edge-on in the sky with a
major-to-minor axis ratio $a/b>9$. Unlike typical disk galaxies,
superthin galaxies often host weak or no galactic
bulges. Spectroscopic studies of individual galaxies
\citep[e.g.][]{goad1981, matthews1999, abe1999, vdkruit2001,
  kurapati2018} have shown that these objects tend to have low
metallicities, low star formation rates (SFRs), as well as slowly
rising rotation curves indicating modest central concentrations. There
have also been efforts to construct specific catalogues for
statistical studies of superthin galaxies
\citep[e.g.][]{karachentsev1993, karachentsev1999, kautsch2006,
  bizyaev2014, bizyaev2017}. Statistically, the superthin galaxies in
the local Universe are preferentially found in low density
environments \citep{kautsch2009, bizyaev2017}. This finding is expected
considering that galaxies in low-density regions have experienced less
dynamical perturbations than those in high-density regions, thus are
easier to maintain a thin disk morphology over a reasonably long
timescale.

Superthin galaxies are selected by smaller ratios between the minor
and major axes, which are respectively relevant to the scale height
and the scale length of a stellar disk. Smaller scale heights and
larger scale lengths naturally lead to superthin disk
morphologies. The scale height is related to a disk's vertical
velocity dispersion (e.g., \citealt{kruit1981}). While the scale
length (size) depends on halo/galaxy angular momentum such that the
larger the angular momentum, the larger the scale length, and the more
rotation-dominant it is for the disk kinematics (e.g.,
\citealt{fall1980, mo1998, Rodriguez-Gomez2017}). It has been known
that gas accretion and angular momentum buildup play important roles
in shaping a galaxy's morphology and kinematics. Traditional theory of
galaxy formation and evolution combines gravitational processes,
dominated by dark matter, dissipative gas dynamics, and further star
formation (e.g., \citealt{white1978, fall1980, blumenthal1984,
  mo1998}). In this ideal scenario, dark matter collapses into
gravitationally-bound halos at early epochs, and gas condenses in the
potential well of these dark matter halos. Stars form from such
condensed gas cores, and appear as a disk if rotation
supported. During this process, angular momentum plays an important
role. Both dark matter and gas acquire initial angular momenta through
the tidal torques mechanism (\citealt{peebles1969, doroshkevich1970,
  white1984, barnes1987}) during the linear evolution
stage of an over-dense region. After turnaround, dark matter halos
collapse, gas falls in along while cooling. Assuming conservation of
specific angular momentum, rotationally supported gaseous disks may
form. Further through star formation, part of the gas angular momentum
transfers to stars and flat rotating stellar disks eventually grow.

However, linear growth and spherical collapse of well-mixed gas and
dark matter are over-simplified scenarios for galaxy formation. One
shall also take into account processes such as dark-matter halo
mergers, misalignment gas accretion and various baryonic physics
processes during the non-linear evolution stage. All these may
significantly change a galaxy's size, morphology and kinematic status
etc. For example, early cosmological simulations of galaxy formation
already revealed that without energy and momentum feedback to gas, gas
may cool too quickly, resulting in too dense cores and disk galaxies
with too small sizes, inconsistent with observations. To solve this
angular momentum problem, feedback processes shall be needed to
suppress gas cooling in order to form realistic disks
(e.g. \citealt{weil1998, sommerlarsen1999, thacker2000}).

Galaxy merger also plays a key role. In term of galaxy angular
momentum, gas can lose a significant fraction of its specific angular
momentum due to gravitational torques, and flow inwards triggering
central star formation and building up bulges during major merger
processes (e.g. \citealt{barnes1991, navarro1991,
  barnes1996}). However gas can also gain specific angular momentum
due to misalignment in spin between gas and dark matter already at
early epochs (\citealt{vdbosch2002, sales2012}). Latest cosmological hydrodynamical
simulations (e.g., \citealt{zjupa2017}) confirmed that the
increase of gas specific angular momentum is a combined result of
removal of lower angular momentum gas due to feedback, and transfer of
specific angular momentum from dark matter to gas during hierarchical
halo assembly.

In terms of galaxy morphology, all types of merger remnants could be made possible according to different merging conditions such as mass ratios, gas fractions and orbit/spin orientations. For example, major mergers (with larger mass ratios, i.e., merging galaxies having comparable masses) are able to destroy stellar disks, resulting in elliptical galaxies that once host bright quasars in a `blowout' phase shortly after major mergers (e.g., \citealt{toomre1972, toomre1977, robertson2006}). However, disk-dominant galaxies may survive or reform even after intense (major) mergers in extreme cases of high gas fractions (\citealt{springel2005apj, athanassoula2016, Rodriguez-Gomez2017}) or with prograde coplanar merging orbits (\citealt{zeng2021, lu2022CQ}). In addition, smaller mass-ratio mergers can also increase disk thickness and enhance bulge growth through repeated bombardments by incoming satellite galaxies \citep{quinn1993, toth1992, walker1996, Purcell2010}. In extreme cases, successive minor mergers could even transform disk galaxies to spheroidals and ellipticals (e.g., \citealt{bournaud2007, Rodriguez-Gomez2017}). The exact morphological outcome due to disk heating may depend on merging orbital parameters (e.g., retrograde orbits, radial orbits, and higher orbital inclination angles of incoming satellites tend to reduce disk heating, see \citealt{velazquez1999, hopkins2009, purcell2009}), as well as mass fractions in gas, which also helps to stabilize the disk and thus prevents disk heating and thickening (e.g., \citealt{purcell2009, moster2010}).

In our recent work (Hu et al. in preparation), we constructed an observational sample of superthin galaxies. We measured shapes of these galaxies at multiple wavelengths, ranging from optical to near-infrared. We found that both the scale length and scale height decrease from optical to near-infrared, however the ratios between them show no significant change, opposite to the result of \cite{bizyaev2020}. We note that observing superthin disks in the near-infrared band suggests that such morphologies also hold for older stellar populations in the systems. This indicates that these galaxies must have suffered less from disk heating (we refer reader to the discussion of our paper). As can be seen, a superthin disk morphology shall require mechanisms that can constructively contribute to growth of gas and disk angular momenta, as well as effectively suppress disk heating, in order to form morphologically extended and rotationally-supported stellar disks. In this study, we investigated possible formation mechanisms of superthin galaxies using a cosmological hydrodynamical simulation. We traced the merger history of a selected sample of superthin galaxies at $z=0$ and studied the statistical properties of mergers they undergo. Through comparisons to a carefully constructed control galaxy sample, we found that an increasing amount of low mass-ratio prograde merging activities since $z\sim 1$ have played a key role in the formation and maintenance of the superthin disk morphologies. 

This paper is organized as follows: In \secref{sec:data}, we introduce briefly the cosmological hydrodynamic simulation used in this work and then describe the construction of a superthin disk sample, and a control sample, both from the simulation in \secref{subsec:samp}. We present the merger tree construction and definitions of a few key merging parameters in \secref{subsec:props_merger}. A number of galaxy properties are traced and we present their detailed definitions in \secref{subsec:trace_propt}. 
In \secref{sec:result}, we firstly present redshift evolution of galaxy shapes in \secref{subsec:evol_ba} and the statistical results of merger properties in \secref{subsec:mergstat}. We investigate key property evolution of (central) galaxies in the context of merger in \secref{subsec:gal_evolv}.
Halo environment dependence and formation of satellite superthin galaxies are presented and discussed in \secref{subsec:cent_satel}. In the end of the paper, conclusions and discussion are given in \secref{sec:conclu_discu}.

\section{Methodology}
\label{sec:data}

\subsection{IllustrisTNG and Sample Selection}
\label{subsec:samp}

In this work, we used simulated galaxies from The Next Generation Illustris simulation suite (IllustrisTNG; \citealt{marinacci2018, naiman2018, nelson2018, pillepich2018b, springel2018}), which is a series of cosmological magnetohydrodynamical simulations with a range of volume and resolution, carried out with the moving-mesh code {\tt AREPO} \citep{springel2010} and a comprehensive model for galaxy formation and evolution \citep{weiberger2017, pillepich2018a}. Our work is based on IllustrisTNG-100 simulation (hereafter TNG100), run within a cubic box of 110.7 Mpc side length and with a mass resolution of $1.4\times10^6{\rm M_\odot}$ and $7.5\times10^6{\rm M_\odot}$, respectively for the baryonic and dark matter. A total of 100 snapshots have been stored across cosmic time. In each snapshot galaxies in host dark matter halos are identified with the {\tt Subfind} algorithm \citep{springel2001, dolag2009}. Properties of galaxies are calculated and publicly released by the IllustrisTNG collaboration \citep{nelson2019}.

\begin{figure}
	\includegraphics[width=\columnwidth]{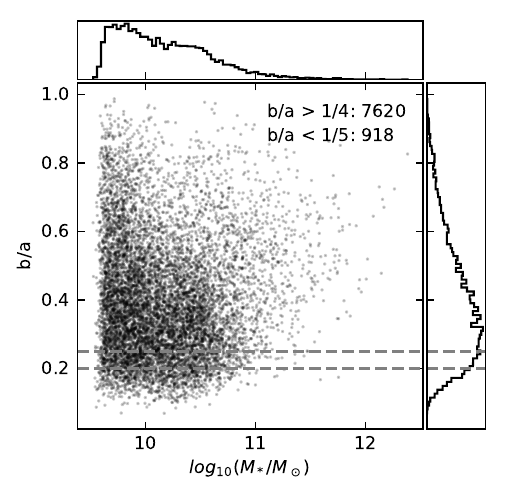}
	\caption{Edge-on axis ratio $b/a$ as a function of stellar mass $M_\ast$. The stellar mass here is calculated within $2 R_{\rm hsm}$. The top and right small panels show histograms of the corresponding parameters. Two grey dashed horizontal lines show criteria for selecting the superthin and the control sample.}
	\label{fig:ba_mass_tot}
\end{figure}

We started by constructing a parent sample of all galaxies at $z=0$ that have stellar masses greater than $5 \times 10^9 {\rm M_\odot}$, defined to be the total mass of star particles within central 30 kpc of the galaxy. This mass cut guarantees sufficient number of stellar particles ($\ga 3000$) in a galaxy. We selected superthin galaxies from the parent sample according to their axis ratios as observed from edge-on.  This was achieved in three successive steps. Firstly, a 3D shape of a given galaxy as described by a triaxial ellipsoid is determined from the stellar mass-weighted inertial tensor, following the method described in \citet{allgood2006} and using stellar particles within a radius of $\min(30\,{\rm kpc},\,3 R_{\rm hsm})$ from a galaxy centre, where $R_{\rm hsm}$ is the radius enclosing half of the total stellar mass. Next, for each galaxy the edge-on view was obtained by projecting all particles onto the plane formed by the longest and the shortest axes according to the rotation matrix obtained in the first step. In the simulation, the raw luminosities of stellar particles were calculated using stellar population synthesis model {\sc galaxev} \citep{bruzual2003} assuming a \citet{chabrier2003} initial mass function. Here we followed the same procedure as in \citet{dandan2017} to implement a simple semi-analytical approach to deal with dust attenuation of the optical lights of stellar particles. Finally, the `observed' minor-to-major axis ratio ($b/a$) was then obtained in SDSS-$r$ band through the luminosity-weighted second moments of the projected luminosity distribution in the edge-on view of a given galaxy (see Section 2.2 of \citet{dandan2017} for details).

\figref{fig:ba_mass_tot} displays the distribution of all galaxies in the parent sample on the plane of $b/a$ versus $\log_{10}M_\ast$. The stellar mass $M_{\ast}$ here is calculated within $2 R_{\rm hsm}$.  We selected superthin galaxies from the parent sample by the criterion of $b/a < 1/5$, which is a relatively loose requirement compared to the definition adopted in observational studies ($b/a < 1/9$). As can be seen from \figref{fig:ba_mass_tot}, the distribution of $b/a$ for superthin galaxies declines sharply at the higher-mass end, indicating that superthin galaxies are rare and a substantially large parent sample is needed if one were to have a significant sample of galaxies with extremely small $b/a$. The limited number of superthin galaxies in the simulation is due to the limited volume and resolution of the simulation. We note that one shall be cautious when comparing the results of this work with any observational studies.

\begin{figure*}
    \includegraphics[width=\textwidth]{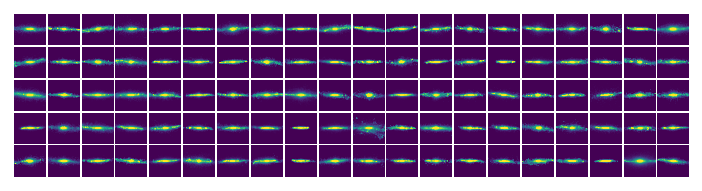}
    \includegraphics[width=\textwidth]{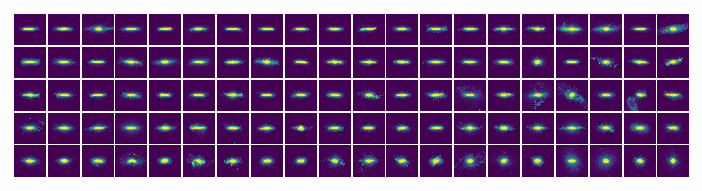}
    \caption{Demonstration of edge-on view of some examples from two samples. Image is for r-band luminosity of galaxies. In each image, the semi-width corresponds to physical length of $5\times {\rm min}(10~{\rm kpc},~R_{\rm hsm})$ and horizontal direction is for longest principal axis determined in 3D space. Top panel is first 100 low b/a galaxies from superthin sample. Bottom panel is 100 galaxies randomly chosen from control sample. Images in both panels are sorted by b/a from top left to bottom right.}
    \label{fig:demo_proj_img}
\end{figure*}

For comparison, we constructed a control sample of galaxies with $b/a > 1/4$ that were also closely matched to the superthin galaxy sample in both $M_\ast$ and the SDSS $g-r$ color. To do so, we first visually examined all galaxies in the superthin sample and excluded six galaxies which are not in flat shapes. Secondly, both the full sample of galaxies with $b/a>1/4$ and the superthin galaxy sample with $b/a<1/5$ were re-sampled so that the two samples shall have the same joint probability density distributions in $M_\ast$ and $g-r$.
We note that some relatively massive superthin galaxies were dropped due to no counterparts in the control sample.
Our final samples include 380 superthin galaxies (312 central galaxies and 68 satellite galaxies) and 3148 control galaxies (2186 central galaxies and 962 satellite galaxies). \figref{fig:demo_proj_img} displays the edge-on images for some example galaxies from both the superthin sample and the control sample. 

\begin{figure}
    \includegraphics[width=\columnwidth]{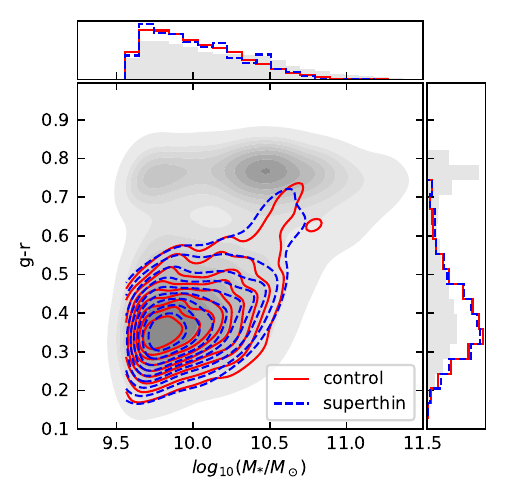}
    \caption{Distribution of samples in color-mass diagram. Grey
      background is for initial sample of all galaxies which have stellar masses $M_* \geq 5 \times 10^9 M_\odot$ within central 30 kpc (all grey dots in \figref{fig:ba_mass_tot}). Dashed blue contours are for galaxies with $b/a < 1/5$. Blue dots are final superthin sample used in statistics study. Red contours are for
      control sample. Top and right panels are margin distributions in corresponding parameter with line color and style coded for different samples same as the central diagram.}
    \label{fig:dist_samples}
\end{figure}

\begin{figure}
	\includegraphics[width=\columnwidth]{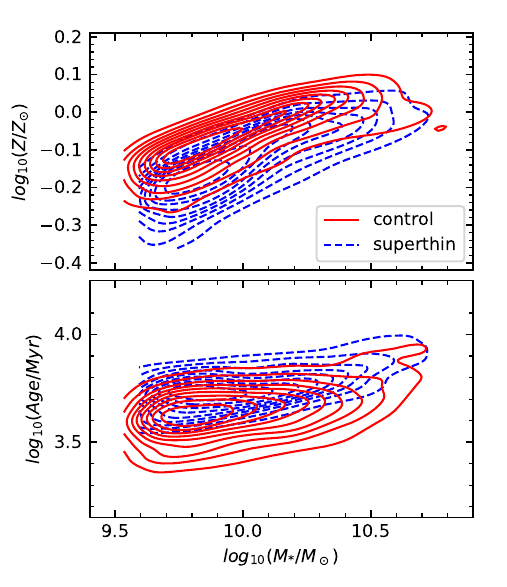}
\caption{Upper panel: distributions of stellar-mass weighted metallicity $\log_{10}(Z/Z_{\odot})$ as a function of stellar mass $M_{\ast}$ for both superthin and control sample galaxies. Lower panel: distributions of stellar age $\log_{10}({\rm age/Myr})$ as a function of stellar mass $M_{\ast}$ for both galaxy samples. Both metallicity and age properties were calculated using stellar particles projected within $R_{\rm hsm}$ along the face-on directions of galaxies. }
\label{fig:metal_age}
\end{figure}

\figref{fig:dist_samples} displays the color-mass
diagram for the final samples, with superthin galaxies plotted in blue
contours and the control sample in red contours.
As can be seen, the final samples are dominated by relatively low-mass galaxies ($M_\ast \lesssim 5\times 10^{10} {\rm M_\odot}$) with bluer colors ($g-r\lesssim 0.7$). We note that there are about twenty superthin galaxies that are relatively more massive and have redder colors. Their presence accounts for the right shoulder of the contours in the diagram. We caution the reader that the conclusion drawn in this study may not apply to this minority ($<10\%$) as outliers of the populations that we investigate in this work.
In \figref{fig:metal_age} we present the distributions of stellar metallicity $\log_{10}(Z/Z_{\odot})$ and stellar age $\log_{10}({\rm age/Myr})$ as a function of stellar mass $M_{\ast}$ for both galaxy samples. Both metallicity and age were calculated using stellar particles projected within $R_{\rm hsm}$ along the face-on direction. We note that, by construction, the two galaxies samples have similar colors (and we also verified that the two samples have similar {\it total} SFRs). However, their central stellar populations differ in stellar age and metallicity. Superthin galaxies host less metal-rich and older stellar populations in their central regions than the control sample, consistent with observations. 

We note the reader that in the Appendix \ref{appd:props}, we particularly provide further comparisons among {\it central} galaxies on a few other basic galaxy properties between the two galaxy samples (in \figref{fig:props_sup_cont}), as well as between a superthin sample (i.e., with $b/a < 0.2$) and a non-superthin ($b/a > 0.25$) sample that were selected not in a controlled fashion by requiring the same joint probability distribution, but merely by asking for the same occupation in the $M_{\ast}$-color space (in \figref{fig:props_sup_nonsup}). As can be seen, general superthin galaxies on average have larger stellar and halo masses and have larger halo spins and stellar radii, comparing to non-superthin galaxies. In particular, among {\it central} galaxies that have stellar masses within $ 9.7 \leqslant \log_{10}(M_*/M_\odot) \leqslant 10.7$ and $g-r$ colors less than 0.7, the fractions of superthin galaxies (i.e., $b/a < 0.2$) are 12\%, 21\% and 24\% in the logarithmic stellar mass bins $\log_{10}(M_*/M_\odot)$ of $9.7 \text{--} 10$, $10 \text{--} 10.3$ and $10.3 \text{--} 10.7$, respectively.

\subsection{Merging History}
\label{subsec:props_merger}

The merger tree dataset we used has been constructed based on dark
matter particles at the subhalo level using the {\tt SubLink}
algorithm \citep{rodriguez-gomez2015}. For each subhalo (containing a
galaxy), a unique descendant in the following snapshot is
assigned. The descendant shall have common particles as its ancestors
and shall have the highest score according to a merit function that
links to the binding energy rank of (dark matter) particles. Then the
merger tree is hierarchically built up based on these descendant
connections. Progenitors of a subhalo are arranged by a mass history
scheme \citep{delucia2007}. In particular, the main/first progenitor
is chosen as the one with the `most massive history', of which the
branch contains most of the mass in a final system for the longest
period. The next/second progenitor is the one with the next most
massive history. The merging tree dataset therefore contains a
collection of merger events that subhalos have ever suffered in the
past. 

With the merger tree, detailed conditions for any given merging event can be
calculated using the main/first and next/secondary progenitors of a descendant
subhalo. In this study, we define several quantities with which we give key
descriptions to the merging events. First of all, in order to identify
merger epochs, we considered two time stamps for any given merging
event:
\begin{itemize}
\item[] $t_{\rm desc}$: time of the snapshot where the descandant subhalo
  (that lies in the main progenitor branch) has more than one
  progenitors in the adjacently previous snapshot. This is also taken
  as the end of a merging event.
\item[] $t_{\rm max}$: time of the snapshot where the secondary progenitor
  reaches its maximum stellar mass along {\it its own} main progenitor
  branch. This is taken as the beginning of a merger, after which the
  secondary progenitor may begin to lose its stellar mass as it starts
  to interact with and be accreted by the main progenitor.
\end{itemize}

With these time stamps identified, we can calculate the frequency of
merging events that a galaxy has experienced in its history (i.e., of
the main progenitor branch). We note that in this work, such a
calculation was performed to a galaxy sample (being superthin or
control), instead of on a individual galaxy basis. We denote ${\rm
  d}N/{\rm d}t$ to be a mean merger rate for a given galaxy sample. To
measure this quantity as a function of redshift/time, we first
split redshifts into several bins, counted the total number of
merging events that come from all progenitors of galaxies in a given
sample and that fall into each redshift bin, and then divided the
number counts first by the time interval and then by the total number
of galaxies in the sample. Finally we assumed a Poisson noise for the
uncertainty of the mean merger rate. In particular, we used both
$t_{\rm desc}$, $t_{\rm max}$ as the time stamp to label and count
merging events. As can be seen in later sections, our findings do not
change between taking different time stamps for the merger rate
calculation.

We also define three quantities to describe the merging condition at
time of $t_{max}$ (where both progenitors exist and the secondary
progenitor reaches the peak in its stellar mass, just before
infall). These quantities are merging mass ratio $\mu_*$, gas fraction
$f_{\rm gas}$ and orbital angle $\theta_{\rm orb}$. The merging mass
raio $\mu_*$ is defined between the first and secondary progenitors as
follows:
\begin{equation}
  \label{eq:mustar}
  \mu_* \equiv \frac{M_{2, *}}{M_{1, *}},
\end{equation}
where subscripts 1 and 2 represent the first and secondary progenitors, respectively; $M_{i, *}$ is the stellar mass measured within $2 R_{\rm hsm}$ of the corresponding subhalo. We note that in the case where $M_{2, *} > M_{1, *}$ (which may happen since the main progenitor is chosen as the one with the `most massive history', but not the most massive one), $M_{1,*}/M_{2,*}$ is used in order to guarantee $\mu_* \leq 1$. We also note that we only counted mergers with secondary progenitor's stellar mass $M_* > 0$, since subhalos without stellar particles have poor resolution and may be artificial object raised from algorithm. 

We note the reader that in this study, we refer to mergers with stellar mass ratio $\mu_* < 0.01$ as `mini mergers' and those with $\mu_* > 0.01$ as `major and minor mergers'.  As shown in \citet{lu2022CQ}, 0.01 is a typical value splitting the bimodal distribution in $\mu_*$ over all merger events of TNG100 galaxies (see Figure 9 therein).

The merging gas mass fraction $f_{\rm gas}$ is defined as follows:
\begin{equation}
  \label{eq:fgas}
  f_{\rm \rm gas} \equiv \frac{M_{1, \rm gas}+M_{2, \rm gas}}{M_{1, \rm
      baryon}+M_{2, \rm baryon}},
\end{equation}
where $M_{i,\,\rm gas}$ and $M_{i,\,\rm baryon} (\equiv M_{i,\,\rm
  gas} + M_{i, *})$ are the gas mass and the sum of gas and stellar
masses, respectively, both calculated within $2 R_{\rm hsm}$ of the
corresponding subhalo.

The merging orbital angle $\theta_{\rm orb}$ is the angle between the
orbital angular momentum of the secondary progenitor and the spin of
the first progenitor, defined as:
\begin{equation}
  \label{eq:angorb}
  \theta_{\rm orb} \equiv \arccos(\frac{\textbf{j}_{1,\rm spin} \cdot
    \textbf{j}_{2,\rm orb}}{|\textbf{j}_{1,\rm spin}|
    |\textbf{j}_{2,\rm orb}|})
\end{equation}
where $\textbf{j}_{1,\rm spin}$ is angular momentum vector of the
first progenitor subhalo, and $\textbf{j}_{2,\rm orb} \equiv
(\textbf{r}_2 - \textbf{r}_1) \times (\textbf{v}_2 - \textbf{v}_1)$ is
orbital angular momentum vector of the secondary subhalo with respect
to the first one, with $\textbf{r}_i$, $\textbf{v}_i$ being
the position and velocity vectors, respectively, of progenitor $i$.
We note that mergers with $\theta_{\rm orb} \sim 0^\circ$ or $180^\circ$ have coplanar orbits with lower  inclination angles, those with $\theta_{\rm orb} \sim 90^\circ$ have polar orbits with higher inclination angles. While mergers with $\theta_{\rm orb} \ll 90^\circ$ ($\gg 90^\circ$) are referred to as in `prograde' (`retrograde') orbits, i.e., the orbital angular momentum vector is in the same (opposite) direction as the main subhalo spin vector.

We note that all three parameters above are calculated using the two
progenitors in snapshot at $t_{\rm max}$. In some cases where the
corresponding snapshot is missing in the main progenitor branch (such
cases can exist due to the the merger tree algorithm), we used the
mean between snapshots before and after $t_{\rm max}$. This happens
rather rarely and has almost no influence to our final results.  As
shall be seen in later sections, for each galaxy sample we also
counted the mean merger rates for events that meet specific merging
conditions according to $\mu_*$, $f_{\rm gas}$, and $\theta_{\rm
  orb}$. The results are presented in the next section.

\subsection{Traced Galaxy Properties}
\label{subsec:trace_propt}

Once the merger tree is known, the evolution history of a subhalo/galaxy can be studied by tracking along the main progenitor branch. In this study, we traced several galaxy properties for both the superthin and the control sample. These properties include the minor-to-major axis ratio $b/a$, the central star formation rate SFR (measured within $R_{\rm hsm}$), subhalo spin parameter $\lambda'_{\rm subhalo}$ and ex-situ stellar mass fraction $f_\exsitu{}$. In particular, the latter two quantities are important parameters related to angular momentum buildup and galaxy merger (e.g., \citealt{Rodriguez-Gomez2017}). We give detailed definitions here below. 

Subhalo spin parameter $\lambda'_{\rm subhalo}$ was defined according to \citet{bullock2001}:
\begin{equation} 
\label{eq:hspin}
\lambda'_{\rm subhalo} \equiv \frac{J_{200}}{\sqrt{2} M_{200} V_{200} R_{200}}.
\end{equation}
In the original definition, $R_{200}$ is the radius of a sphere enclosing a mean density equal to 200 times the critical density of the Universe. Here we directly took `$\rm Group\_R\_Crit200$' in the FoF Group Catalogue for an approximation. $M_{200}$ in principle should be the total mass within $R_{200}$; here we used the property `SubhaloMass' from the Subhalo Catalogue for an approximation. We note that for central galaxies, these can be treated as good approximations. $V_{200}=\sqrt{G M_{200} / R_{200}}$, with $G$ being the gravitational constant. $J_{200}$ is the total angular momentum calculated using all particles within $R_{200}$ of a subhalo. Here we used `SubhaloSpin' from the Subhalo Catalogue for an approximation of $J_{200}/M_{200}$.

The \exsitu{} stellar mass fraction $f_\exsitu{}$ is obtained from the TNG Supplementary Data Catalog for stellar assembly \citep{rodriguez-gomez2015, rodriguez-gomez2016, Rodriguez-Gomez2017}. In brief, each stellar particle is traced to the very source where it was born. If the source is along the main progenitor branch, it is counted as \insitu{}; otherwise as \exsitu{}. This parameter essentially counts the mass fraction of stars that were born elsewhere and accreted to the current galaxy later on. In general, the higher $f_\exsitu{}$ is, the more frequent dry merging events (i.e. merging with low gas fraction) a subhalo has experienced. 

Examining the evolution of these galaxies properties, along with the merger history, can further help us better understand the formation of present-day superthin morphologies in the context of galaxy merger. We present the results in the next section.

\section{Result}
\label{sec:result}

\subsection{Shape Evolution}
\label{subsec:evol_ba}

\begin{figure}
    \includegraphics[width=\columnwidth]{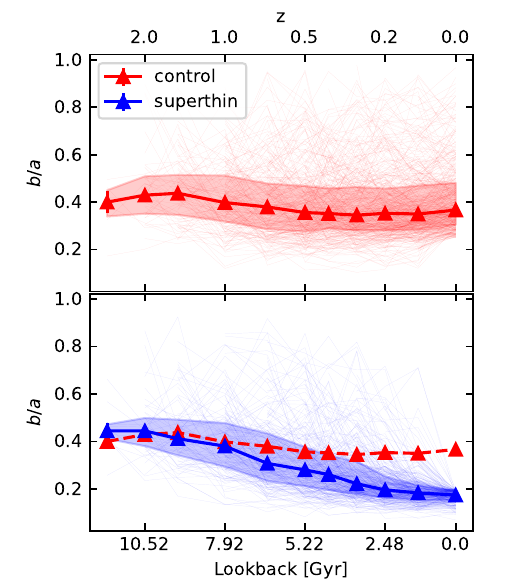}
    \caption{Evolution of edge-on axis ratio $b/a$ as a function of lookback (labelled on the bottom) and redshift (labelled on top). Each thin curve in the background represents the evolution of a given galaxy. Triangles mark the median values of $b/a$ of progenitors calculated at major snapshots in the simulation. The shaded regions present the interquartile range of [0.25, 0.75] in $b/a$. The top panel is for the control sample, while the bottom panel is for the superthin sample, where the red dashed line is a copy of the median curve of the control sample presented in the top panel.}
    \label{fig:ba_evol}
\end{figure}

The first question we address is whether the extreme shape 
of superthin galaxies is formed at birth or develops with time. 
To answer this question, for each galaxy in our samples as selected from the 
snapshot at $z=0$, we tracked back in time along its main progenitor 
branch and examined how the minor-to-major axis ratio $b/a$ of the progenitors
evolves with redshift. In \figref{fig:ba_evol}, we show $b/a$ as a function of
redshift (and lookback time)
for both the superthin and the control sample. The thin curves 
plotted in the background are randomly selected galaxies as examples, 
while triangles connected by a solid thick line and the shaded region show the 
median and the interquartile range of $[0.25, 0.75]$ 
of $b/a$ at given redshift. Although on
individual basis galaxies in both samples present marked rise and fall in $b/a$, 
on average the two samples show clear but different evolution trends with redshift.
For superthin galaxies, the evolution of 
$b/a$ can be divided into three stages: a slow decrease at earlier times 
with $b/a\sim0.4$ at $z\ga 1$, a fast decrease phase over $0.2\la z\la 1$
reaching $b/a\sim0.2$ at $z\sim0.2$, and a slow decrease again ending
up with $b/a\lesssim 0.2$ at $z=0$. In contrast, the control sample
behaves similarly to the superthin galaxies only at early times ($z\ga 1$),
but at later times its average $b/a$ has been roughly a constant at 
$b/a\sim0.4$. As a result, the two samples deviate from each other 
at $z\sim1$. This result suggests that the unusually thin shape of the 
superthin galaxies is not formed by nature, but rather, it should be attributed 
to some processes occurring mainly after $z\sim1$ and before $z\sim0.2$. 

\begin{figure*}
    \includegraphics[width=\textwidth]{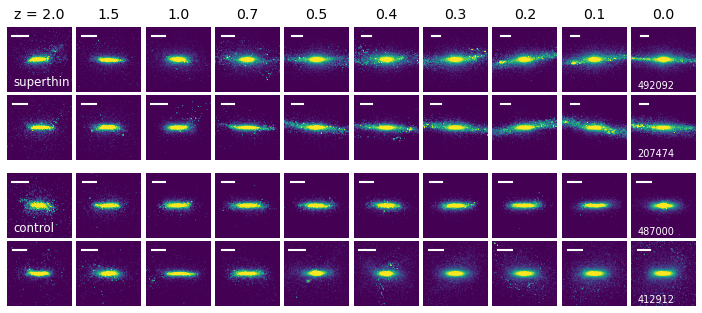}
    \caption{Edge-on view of galaxies' progenitors at a set of redshifts (marked on top of each column) since $z=2$. Each row is for a galaxy, whose {\tt Subfind} ID is marked in right-most panel. Top two rows are for galaxies from the superthin sample and bottom for those from the control sample. In each image, the white line marks a physical length of 5 kpc. The semi-width of each image is $5\times \min(10~ {\rm kpc}, ~R_{\rm hsm})$.}
    \label{fig:demo_ba_evol}
\end{figure*}

\figref{fig:demo_ba_evol} displays how the edge-on view has changed 
with redshift for four example galaxies, including two superthin galaxies 
and two control galaxies. They are chosen to have similar $b/a$ at early
times. Both the flattening procedure of the superthin galaxies
and the steady shape development of the control galaxies can be clearly seen from 
the figure. To guide the eye, a white horizontal bar indicating a fixed physical 
length of 5 kpc is included in each panel. By the shrinking of this line with 
decreasing redshift, one can more clearly see the significant growth in disk 
lengths of superthin galaxies.
In contrast, such extended disk structures are not present for the 
control galaxies, which grow steadily in both disk length and thickness, 
maintaining a similar shape over a long period of time.
We also mention in passing that 
the flattening process of superthin galaxies is mainly the construction of its extended disk structure. During this process, they must have suffered less from disk heating in order to maintain small axis ratios.

\subsection{Statistics of Merging History}
\label{subsec:mergstat}

\begin{figure}
    \includegraphics[width=\columnwidth]{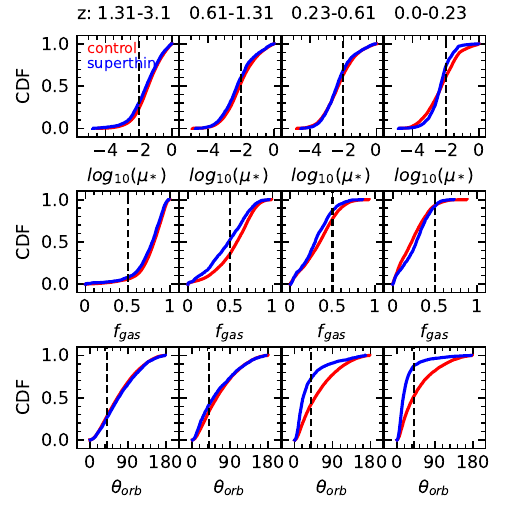}
    \caption{Cumulative distribution functions of merging mass ratio $\log_{10} \mu_*$ (top), gas fraction $f_{\rm gas}$ (middle) and orbit-spin angle $\theta_{\rm orb}$ (bottom). For this plot, we used $t_{\rm desc}$ as time points of a merger to divide it into four different redshift bins, marked on top of the figure. The blue and red lines represent the superthin and control samples, respectively. }
    \label{fig:contour_gas_binz}
\end{figure}

\figref{fig:contour_gas_binz} shows the cumulative distribution functions of merging mass ratio $\log_{10} \mu_*$, gas fraction $f_{\rm gas}$, and orbit-spin angle $\theta_{\rm orb}$ in four different redshift bins. As can be seen, galaxies in general experienced more and more smaller mass-ratio mergers and less and less larger gas-fraction mergers towards lower redshifts. This is an natural consequence of hierarchical galaxy assembly and cosmological gas consumption. When comparing the two galaxy samples, superthin galaxies experienced slightly less frequent larger mass-ratio ($\mu_*>0.01$) mergers at $z\la 0.2$, and more frequent lower gas-fraction mergers at at $0.2<z<1$ than their control galaxy sample. But the most established difference between the two samples lies in the merging orbital angle. The two samples have same distributions in $\theta_{\rm orb}$ at higher redshifts but develop divergence at below $z\sim 1$. In particular, at $z<0.6$ superthin galaxies have experienced notably more small-angle, i.e., prograde merging events than galaxies in the control sample. 

It is worth noting that gas fraction and merging orbit play vital roles in the morphological outcome of a merging event. On the one hand, from disk heating perspective, previous theoretical studies (see Section \ref{sec:intr}) have shown that higher gas fraction, as well as orbits with high inclination angles or retrograde orbits  
are generally favored properties that can effectively suppress disk heating and thickening. Interestingly, these features are not notably present in the merging history of superthin galaxies. On the other hand, existing studies also suggest that in the case of substantially high gas fraction or prograde merging orbit, disk morphologies with rotating kinematics are likely to be re-established or maintained.  In order to understand the role of these two key aspects in the formation of superthin galaxies, we examined detailed redshift evolution under different merging gas fractions and orbital conditions. 

\begin{figure}
    \includegraphics[width=\columnwidth]{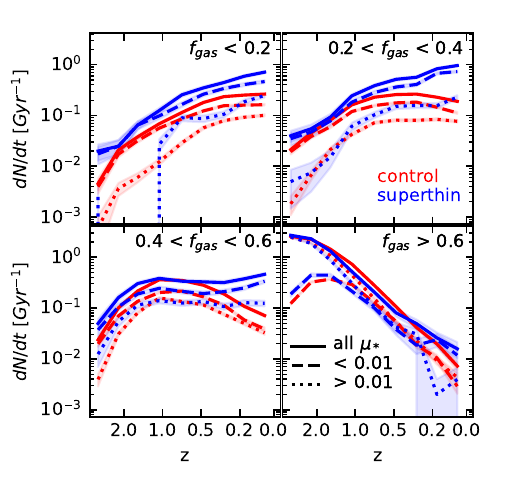}
    \caption{Merger rates ${\rm d}N / {\rm d}t$ of merging events specified by $f_{\rm gas}$ (labelled in four sub-panels) as a function of redshift $z$ for the two galaxy samples. The blue and red lines represent the superthin and control samples, respectively. We also divided mergers by their stellar mass ratio into $\mu_* < 0.01$ (dashed lines) and $\mu_* > 0.01$ (dotted lines). 
    }
    \label{fig:merger_rate_binfgas}
\end{figure}

\figref{fig:merger_rate_binfgas} shows the merger rates in four different $f_{\rm gas}$ bins among the two galaxy samples. We also divided mergers by their stellar mass ratio $\mu_*$. As can be seen here, on the whole both galaxy samples have experienced a dominating number of gas-rich and larger mass-ratio mergers ($f_{\rm gas} > 0.6$, $\mu_* > 0.01$) before $z\sim 1$. After this redshift, superthin galaxies undergo an increasing amount of mergers with $f_{\rm gas} < 0.6$, which become more and more dominated by mini mergers as $f_{\rm gas}$ decreases. In particular, the rates of mergers with $f_{\rm gas} \leqslant 0.4$ show clear increasing trends for superthin galaxies towards lower redshifts, while the rates of mergers with $f_{\rm gas} > 0.6$ continuously decrease with time. It is evident that the control sample galaxies have not experienced as many lower $f_{\rm gas}$ merging events as superthin galaxies towards lower redshifts. One shall notice that the number of mini mergers is several times higher than the number of major mergers (for lower-gas fraction mergers) in these systems. It is worth noting that the differences between the two galaxy samples in rates of mergers with lower gas fractions (as defined in Section \ref{subsec:props_merger}) are essentially addressed to minor mergers, and thus largely dominated by the difference in the gas content in the main progenitors but not lying in the difference of gas brought in by incoming secondary galaxies (this can also be seen from the difference in central SFRs between the two galaxy samples as presented in \figref{fig:history_props_central} of Section \ref{subsec:trace_propt}, which shows that superthin galaxies have notably lower central SFRs than their control sample counterparts).   

\begin{figure}
    \includegraphics[width=\columnwidth]{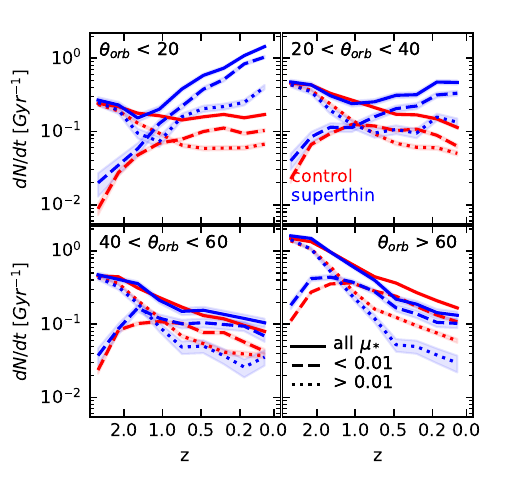}
    \caption{Same as \figref{fig:merger_rate_binfgas}, but merging events specified according to the orbit-spin angle $\theta_{\rm orb}$ (labelled in four sub-panels).}
    \label{fig:merger_rate_binang}
\end{figure}

\figref{fig:merger_rate_binang} shows the merger rates of events specified according to orbit-spin angle $\theta_{\rm orb}$ for the two galaxy samples. As can be seen, for merging events with $\theta_{\rm orb} > 40^{\circ}$, the two samples have similar evolution at all redshifts. However, a significant difference between the superthin and the control sample starts to develop around $z\sim 1$ for the merging events with orbital angle $\theta_{\rm orb} \leqslant 40^{\circ}$: superthin galaxies undergo a significantly increasing amount of prograde-type mergers since. We expect these merging events to have transformed a remarkable fraction of orbital angular momenta to spins of galaxies in the main progenitor branch. We shall see in next sections that this particular merging aspect plays a significant role in developing superthin morphologies. 

\subsection{Key Property Evolution of Central Galaxies in the Context of Merger}
\label{subsec:gal_evolv}

Galaxy merger plays a key role in regulating galaxy angular momentum and shaping the morphology and kinematics of galaxy disk. In Section \ref{subsec:mergstat}, we present detailed merging histories and comparisons between superthin and normal galaxies. To better understand impact of mergers under different conditions, in this section, we present evolution of a number of key galaxy properties that can be mostly affected by merger. We note that this subsection presents results that were only obtained for central galaxies in both samples. 

\begin{figure}
    \includegraphics[width=\columnwidth]{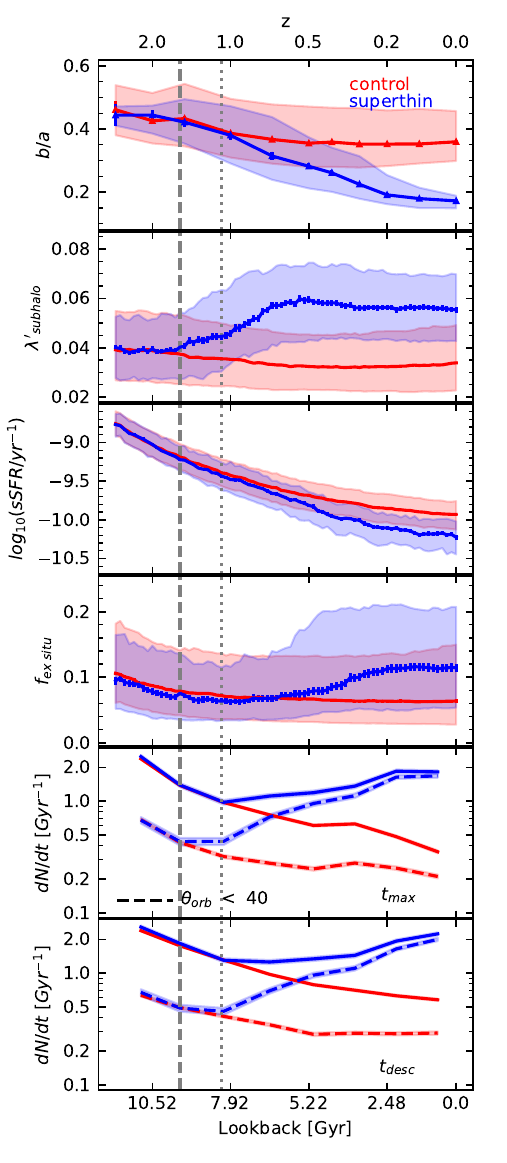}
    \caption{Top four panels: redshift evolution of edge-on axis ratio $b/a$, subhalo spin $\lambda'_{\rm subhalo}$, SFR within $R_{\rm hsm}$, ex-situ fraction $f_\exsitu{}$. Lines and shades represent the median and interquartile range of $[0.25, 0.75]$ at each redshift. Bottom two panels: merger rates ${\rm d}N/{\rm d}t$ counted under different orbital conditions and using different time stamps (i.e., $t_{\rm max}$ and $t_{\rm desc}$, labelled in the panel). The dashed lines indicate the merger rates for prograde mergers with $\theta_{\rm orb} \leqslant 40^\circ$. Poisson noise is adopted as the uncertainty for the calculated mean merger rate (see Section~\ref{subsec:props_merger}). The blue and red color represent the superthin and control samples, respectively. Grey vertical lines indicates two key timelines (see main text for details).}
    \label{fig:history_props_central}
\end{figure}

The top four panels in \figref{fig:history_props_central} present the redshift evolution of (edge-on) axis ratio $b/a$, subhalo spin $\lambda'_{\rm subhalo}$, central SFR within $R_{\rm hsm}$, ex-situ fraction $f_\exsitu{}$ (see Section~\ref{subsec:gal_evolv} for detailed definitions). Lines and shades represent the median and interquartile range of $[0.25, 0.75]$ at each redshift. In the bottom two panels, we present the merger rates counted under different orbital conditions and using different time stamps (i.e., $t_{\rm max}$ and $t_{\rm desc}$, considered as the start and the end time point of a merger). In order to highlight the significance of prograde mergers to superthin galaxies (as demonstrated in \figref{fig:merger_rate_binang}), we counted the merger rates for prograde mergers with $\theta_{\rm orb} \leqslant 40^\circ$, which are plotted separately by the dashed lines. For the merger rates, we assumed a Poisson noise for the uncertainty in the calculated mean merger rates (see Section~\ref{subsec:props_merger}). 

As can be seen, superthin galaxies show significant divergence from the control sample galaxies after some time points. Two noticeable diverging points in time are marked by the grey vertical lines. The dashed vertical line at $z\sim 1.5$ marks the diverging point of the mean ${\rm d}N/{\rm d}t$ that was calculated using the starting points $t_{\rm max}$ of merging events with $\theta_{\rm orb} \leqslant 40^\circ$, after which superthin galaxies have experienced an ever increasing number of prograde mergers. We notice that around similar redshift $\lambda'_{\rm subhalo}$ of the two populations also diverge: subhalo spins of superthin galaxies climb up quickly, from a median value of 0.04 at $z\sim 1.5$ to 0.06 at $z\sim 0.7$. This reflects that prograde merging events (already at halo incoming stages) may have contributed positively to the buildup of dark matter halos' angular momenta. 

While the dotted vertical line at $z\sim 1$ marks the time of significant divergence between the two populations in ${\rm d}N/{\rm d}t$ that was calculated using the end points $t_{\rm desc}$ of merging events with $\theta_{\rm orb} \leqslant 40^\circ$. Again we also notice that roughly around/soon after this redshift the average axis ratios $b/a$s of the two galaxy populations also diverge, with superthin $b/a$ ever decreasing from a median value of $\sim 0.4$ at $z\sim 1$ to below 0.2 at $z=0$. This reflects that the final impact of prograde mergers on the shape of the central stellar disk eventually takes place around/after the end of a merging event. It is worth noting that this result strongly echos that of \citet{lu2022CQ}, which investigated the different formation paths of dynamically cold and hot galaxies at the more massive end in the TNG100 simulation. They also found that a dominant number of prograde mergers since $z\sim 1$ plays a key role in maintaining dynamically cold disc morphologies even among quenched galaxy populations (see Figure 8 and 9 therein).

We also note that the time interval between the divergence time points for prograde merger rates counted using $t_{\rm max}$ and $t_{\rm desc}$ is roughly 1 Gyr, which is rather long compared to the interval calculated for the total merger rates. This implies relatively slow decay of these prograde orbits. We note that the merging gas fraction is still high at this redshift (as is shown in \figref{fig:contour_gas_binz}), which further helps to settle down the stellar growing disks. 

It is interesting to notice that shortly after subhalo spin $\lambda'_{\rm subhalo}$ of superthin galaxies starts rising up and developing differently from control sample galaxies (second panel), the SFRs measured within $R_{\rm hsm}$ of superthin galaxies also drops below their normal galaxy counterparts (third panel). In particular, by redshift $z=0$, superthin galaxies have lower metallicities and older stellar populations due to slower star forming activities in their central regions (see \figref{fig:props_sup_cont}). It is worth noting that the stellar disk of a superthin galaxy may extend to nearly 5 times $R_{\rm hsm}$ in comparison to a normal disk galaxy (as shown in \figref{fig:demo_proj_img} and \ref{fig:demo_ba_evol}). The difference present in the central SFRs between the two galaxy samples  actually disappears when comparing their total SFRs across all redshifts; in fact the two samples (by construction) have identical distributions in color (and thus SFR) by $z=0$. We attribute lower SFRs in central regions of superthin galaxies to their higher subhalo spins, which may prevent efficient gas infall to fuel central star formation, and rather lead to star formation happening at larger distances from the galaxy centres, as a result of higher gas angular momentum (\citealt{Wang2022TNGEnvironment, Lu2022TNGEnvironment}).    

The fourth panel in the figure presents the evolution of the ex-situ stellar mass fraction $f_\exsitu{}$. The two samples show divergence at around $z\sim 0.6$, where superthin galaxies start having a notably increasing amount of ex-situ stars accreted onto the systems. This is also expected as a consequence of increasing merger rates (and particularly in form of lower gas-fraction mergers, see \figref{fig:merger_rate_binfgas}) experienced by superthin galaxies since $z\sim 1$. 

It is worth noting that at $z=0$ superthin galaxies have a mean merger rate 3-5 times higher than their control galaxy counterparts, and actually host higher fractions of accreted stars, in comparison to the control sample galaxies. However, they have successfully managed to maintain their superthin disk morphologies, largely thank to a greater number of prograde merges in their history. 

\subsection{Superthin Satellite Galaxies}
\label{subsec:cent_satel}

Central and satellite galaxies are distinguished by their positions in the host halos. In the potential well of the host halo, satellite galaxies gradually move inwards on different orbits towards the center. During this process, gas and stars in satellite galaxies could be stripped from their own dark matter halo and accreted to the central galaxy and/or the host dark matter halo. Considering this, central and satellite galaxies may have different evolution modes in terms of their morphologies and kinematics etc. 
It is interesting to ask whether superthin galaxies can live in larger dark matter halos as satellite galaxies? If so, how do their dark halo environments affect their superthin morphologies? In order to address these questions, we investigate the formation of superthin satellite galaxies. 

\begin{figure}
    \includegraphics[width=\columnwidth]{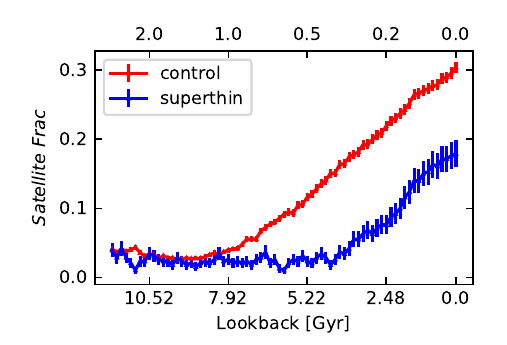}
    \caption{Evolution of satellite fractions of both galaxy samples in history. The blue and red color represent the superthin and control samples, respectively. Poisson noise is assumed as the uncertainty in satellite fraction.}
    \label{fig:history_frac_satel}
\end{figure}

\figref{fig:history_frac_satel} shows the evolution of the satellite fraction in both galaxy samples in history. This fraction is defined as $N_{\rm satellite}/N_{\rm total}$, where $N_{\rm total}$ is the total number of progenitors at a given redshift, and $N_{\rm satellite}$ is number progenitors being satellite galaxies. As can be seen, the two samples show similar (low) satellite fractions before $z\sim 1$. However, a notable difference develops since, and the satellite fraction in the control sample grows much faster since an earlier stage; while the satellite fraction of the superthin sample only starts to increase since $z \sim 0.5$. By $z=0$, the satellite fractions are $18\%$ and $30\%$ for the superthin and the control galaxy sample, respectively. This implies that superthin galaxies fall into their host dark matter halos at a later time.

\begin{figure}
    \includegraphics[width=\columnwidth]{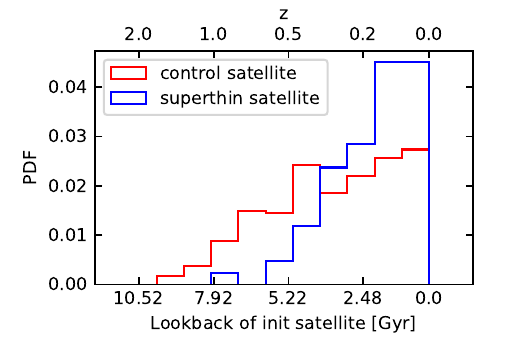}
    \caption{The probability density functions of infall time for satellite galaxies in the superthin sample (blue) and the control sample (red) selected at $z=0$. }
    \label{fig:hist_init_satellite}
\end{figure}

Such a belated infall time for superthin satellite galaxies is confirmed in \figref{fig:hist_init_satellite}, which shows the probability density functions of infall time for satellite galaxies in both samples. The infall time was marked as the time when a present-day satellite first became a satellite (to the final FoF group at $z=0$) in its main progenitor branch.
We shall note the reader that a FoF group in a simulation is not a gravitationally bound system, but merely identified by connecting distances smaller than some threshold value. However, we still use this as some loose definition of host dark matter halos that satellites live in. We expect more severe effects such as tidal and ram pressure stripping happens as satellites get closer and closer towards the centre of FoF group. From the figure, we can see that superthin-sample satellite galaxies fall into their present-day host dark matter halos in much later epochs, hinting possibly much weaker impacts on their superthin morphologies during infalling through the host halo environment, in comparison to their control-sample satellite counterparts.   

\begin{figure}
    \includegraphics[width=\columnwidth]{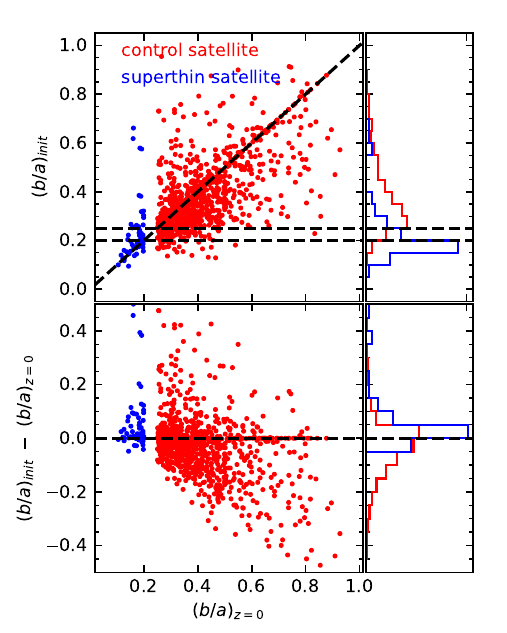}
    \caption{A comparison between the edge-on axis ratio $(b/a)_{\rm init}$ at time of satellite infall and $(b/a)_{z=0}$ at $z=0$ for each satellite galaxy in both galaxy samples (in the top panel). The blue and red color represent the superthin- and the control-sample satellite populations, respectively. The differences are presented in the bottom panel. 
    }
    \label{fig:sca_hist_ba_stallite}
\end{figure}

In order to understand how much the edge-on axis ratio $b/a$ of a present-day satellite has evolved from before it falls into its host halo to its current value, we present \figref{fig:sca_hist_ba_stallite}, which compares $b/a$ at time of satellite infall to that at $z=0$ for each satellite galaxy in both samples. As can be seen, at time of infall almost all superthin satellite galaxies used to have $b/a$ around 0.2 (which is our selection criterion), implying that they already obtained their superthin morphologies by that time. This is not the case for the control-sample satellite galaxies, which already had much thicker morphologies at time of infall. The lower panel shows the difference between the axis ratios at infall and by $z=0$. Superthin satellite galaxies tend to become thinner with time, while control-sample satellites tend to grow thicker after they fall into their host halos. This suggests that present-day superthin satellites galaxies not only have been less affected by various physical processes during infalling through their host halo environment, but also have been able to further develop their superthin morphologies.  

\section{Conclusion and Discussion}
\label{sec:conclu_discu}

In this work, we investigated the formation mechanism of superthin galaxies in the IllustrisTNG simulation. To do so, we selected a superthin galaxy sample which have edge-on axis ratio $b/a < 0.2$ and a control galaxy sample which share the same joint probability distribution in the stellar mass and color plane but have edge-on axis ratio $b/a > 0.25$ (see Section \ref{subsec:samp}). We traced the merger history of both galaxy samples and investigated the impact of mergers under different conditions in terms of merging gas fractions and orbital angles (see Section \ref{subsec:props_merger}). We studied the evolution of several galaxy / halo properties (in morphology, kinematics and star formation) that are mostly affected by merging activities (see Section \ref{subsec:trace_propt}). We found an important role played by galaxy mergers. Our results are as follows: 

\begin{itemize}

    \item At higher redshifts superthin galaxies have same edge-on $b/a$ distribution as the control sample. After $z \sim 1$, $b/a$s of superthin galaxies start to decrease markedly with time, while that of the control sample stays roughly at the same level. A superthin morphology as featured by extremely small $b/a$ is really a consequence of significant growth in disk lengths. By $z=0$, superthin galaxies have disk sizes extending to $\sim5R_{\rm hsm}$, in comparing to $\sim 2R_{\rm hsm}$ for normal disk galaxies (see \figref{fig:demo_ba_evol} and Section \ref{subsec:evol_ba}).

    \item The main reason that causes such diverged growths in edge-on $b/a$ between the two galaxy samples since $z\sim 1$ can be attributed to significantly different prograde merger frequencies experienced by the two populations since such a redshift. While normal galaxies in general undergo less and less frequent mergers since $z\sim 1$, superthin galaxies experience an increasing amount of merging activities towards lower redshifts, which are dominated by mini gas-poor and prograde merging events, with $\mu_* \la 0.01$, $f_{\rm gas} \leqslant 0.4$ and $\theta_{\rm orb} \leqslant 40^\circ$. Such merging activities may have greatly brought angular momenta to the systems, resulting in significant stellar disk growth and thus a reduction in stellar axis ratios towards lower redshifts (see \figref{fig:merger_rate_binfgas}, \ref{fig:merger_rate_binang} and Section \ref{subsec:mergstat}, as well as \figref{fig:history_props_central} and Section \ref{subsec:gal_evolv}).

    \item Towards lower redshifts, {\it central} superthin galaxies also develop lower SFRs in their inner regions than their normal galaxy counterparts. We attribute this to a significant growth of dark matter halo spins of superthin galaxies since $z\sim 1.5$, which can prevent efficient gas infall to the central parts of the systems. In addition, superthin galaxies also develop higher ex-situ stellar fractions since $z\sim 1$, comparing to the control sample counterparts, as a consequence of the overwhelming number of mini and prograde mergers in history of superthin galaxies (see \figref{fig:history_props_central} and Section \ref{subsec:gal_evolv}).

    \item On average, satellite superthin galaxies fall into their host dark matter halos about $2-3$ Gyrs behind their control-sample satellite counterparts (see \figref{fig:history_frac_satel}, \ref{fig:hist_init_satellite}). Superthin satellites already obtained very small $b/a$s before they fall into their host dark matter halos and have maintained their superthin morphologies until $z=0$ (see \figref{fig:sca_hist_ba_stallite}).  

\end{itemize}

These results suggest the following scenario for the formation of superthin galaxies. Current superthin galaxies had normal shape as their control sample counterparts at birth. Their flattening process is triggered by an increasing amount of mini prograde mergers towards lower redshifts since $z \sim 1$ in comparison to their normal disk counterparts. These prograde merging activities increase their dark halo spins and bring in materials with high angular momenta. In particular, newly accreted gas assembles extensively out to large distances from the galaxy centres and the subsequent star formation results in extended stellar disks, which grow even larger with time. 
The predominance of prograde merging orbits since $z\sim 1$ may have also provided a mechanism that prevents efficient gas infall, resulting in smaller central gas fraction and lower central SFRs in superthin galaxies (see also \citealt{Wang2022TNGEnvironment, Lu2022TNGEnvironment}, and Wang et al. in prep). We mention in passing that the effect of disk heating due to the frequent merging activities present in history of superthin galaxies must have not played a significant role in diminishing the superthin morphologies. In particular, the increase of disk scale height due to disk heating must have been relatively slower than that of the disk scale length, in order to  maintain superthin morphologies. 


\appendix

\section{Properties of superthin galaxies at redshift zero}
\label{appd:props}

In this appendix, we present comparisons among $z=0$ {\it central} galaxies on a number of basic properties between superthin galaxies and the rest. For this, we only took central galaxies that have stellar masses $5\times 10^9 \leqslant M_\ast/M_\odot \leqslant 5\times 10^{10} $ and $g-r$ colors less than 0.7, as this is the region in the parameter space where the majority of TNG100 superthin galaxies lie. For each galaxy, both $r$-band magnitude $M_r$ and $g-r$ color were calculated using all stellar particles bounded to the subhalo. Stellar mass $M_\ast$ was calculated using stellar particles within $2 R_{\rm hsm}$. Subhalo mass $M_{\rm subhalo}$ is the total mass of all types of particles. Two specific star formation rates (sSFR) were calculated using stellar particles in different regions: \textit{total} sSFR of the entire subhalo and \textit{central} sSFR within $R_{\rm hsm}$. Subhalo spin parameter $\lambda^{\prime}_{\rm subhalo}$ is defined by Equation \ref{eq:hspin} of \secref{subsec:trace_propt}. The central age and metallicity are the same as \figref{fig:metal_age} in \secref{subsec:samp}.

\begin{figure*}
    \includegraphics[width=\textwidth]{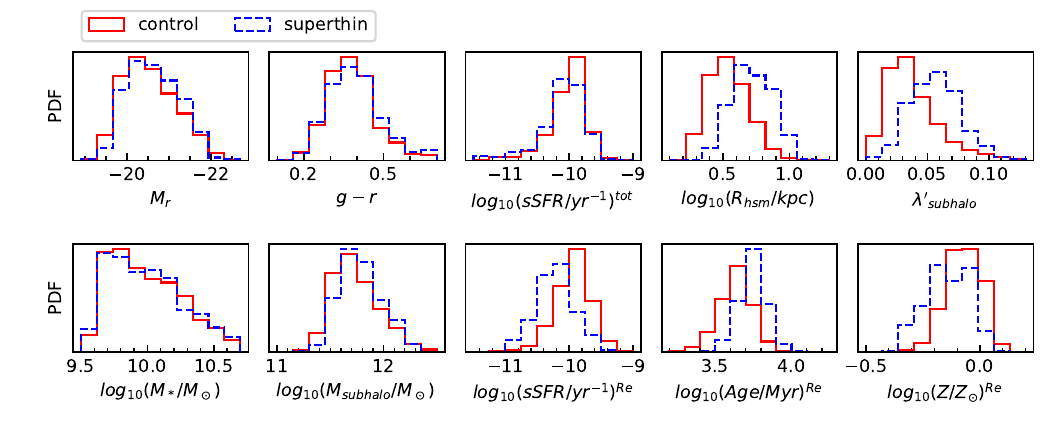}
    \caption{Probability distribution functions of a number of galaxy properties for $z=0$ superthin/control {\it central} galaxies. All galaxies plotted here are required to have stellar masses of $5\times 10^9 \leqslant M_\ast/M_\odot \leqslant 5\times 10^{10} $ and $g-r$ colors less than 0.7. The red solid and blue dashed line represent the control and superthin samples (see Section \secref{subsec:samp} for sample definitions). Panels from top left to bottom right correspond to $r$-band magnitude $M_r$, $g-r$ color, \textit{total} specific star formation rate $\log_{10} \,({\rm sSFR}/[{\rm yr}^{-1}])^{\rm tot}$, half stellar mass radius $\log_{10} (R_{\rm hsm}/{\rm kpc})$, subhalo spin parameter $\lambda^{\prime}_{\rm subhalo}$, stellar mass $\log_{10}(M_\ast/M_\odot)$, subhalo mass $\log_{10} (M_{\rm subhalo}/M_\odot)$, \textit{central} specific star formation rate $\log_{10} \,({\rm sSFR}/[{\rm yr}^{-1}])^{R_{\rm e}}$, central age $\log_{10} ({\rm Age}/{\rm Myr})^{R_{\rm e}}$ and central metallicity $\log_{10}(Z/Z_\odot)^{R_{\rm e}}$.}
    \label{fig:props_sup_cont}
\end{figure*}

In \figref{fig:props_sup_cont}, a comparison between the superthin galaxy sample and the control sample is presented. By construction, the two samples have same distributions in these two controlled properties above. In addition, the two samples also share very similar distributions in their rest-frame $r$-band magnitude $M_r$, total sSFR $\log_{10} \,({\rm sSFR}/[{\rm yr}^{-1}])^{\rm tot}$ and subhalo mass $M_{\rm  subhalo}$. However, it is interesting to note that despite of similar distributions in total sSFR, superthin galaxies have markedly lower \textit{central} sSFR $\log_{10} \,({\rm sSFR}/[{\rm yr}^{-1}])^{R_{\rm e}}$ than their control sample counterparts. As is discussed in \secref{sec:result}, this can be attributed to the high dark matter halo spins $\lambda^{\prime}_{\rm subhalo}$ of superthin galaxies (as can also be seen in the figure), which are built up due to galaxy mergers, and play a key role in preventing efficient gas infall to the central parts of superthin galaxies. As a consequence of such, one shall expect larger sizes in stellar disks as indicated by $R_{\rm hsm}$, as well as older stellar populations and lower metallicities in the central regions of superthin galaxies, in comparison to the control sample galaxies, as are also shown in the figure. 

\begin{figure*}
    \includegraphics[width=\textwidth]{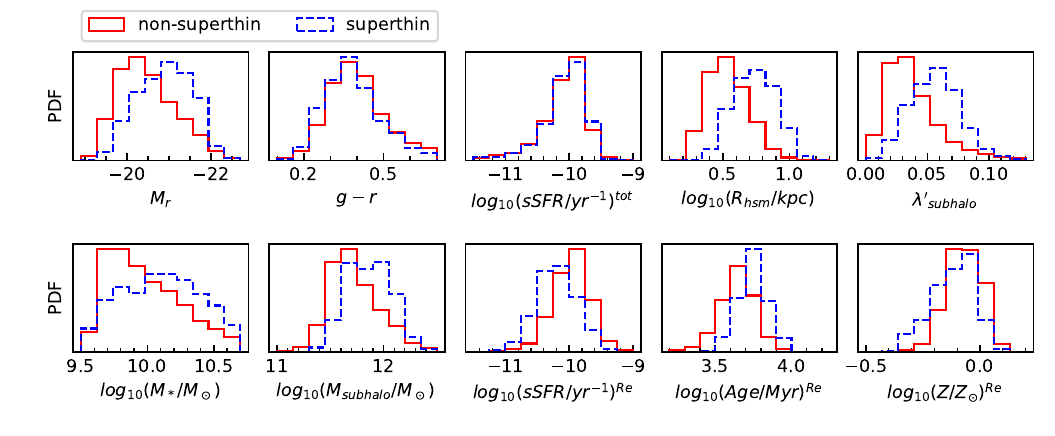}
    \caption{Same as \figref{fig:props_sup_cont}, but the two samples being general superthin galaxies (i.e.,  with $b/a < 0.2$, blue dashed) and non-superthin galaxies (i.e., with $b/a > 0.25$, red solid).}
    \label{fig:props_sup_nonsup}
\end{figure*}

We also investigate the comparison between superthin galaxies ($b/a < 0.2$) and a non-superthin sample ($b/a > 0.25$) that were selected not in a controlled fashion by requiring the same joint probability distribution, but by merely asking for the same occupation in the $M_{\ast}$-color space. The result is presented in \figref{fig:props_sup_nonsup}.
As can be seen, apart from above-mentioned systematic differences that remain evident here, general superthin galaxies on average have larger subhalo masses, larger stellar masses and higher luminosities.

\section*{Acknowledgements}

We acknowledge Sen Wang for useful discussion and proofread of the paper, and Dr. Shengdong Lu for his kind help with the simulation data. We would also like to thank the referee for his very constructive comments that improve the quality of this paper. This work is supported by the National Key Research Development Program of China (grant No. 2022YFA1602902 and 2022YFA1602903). 
\bibliography{references}

\end{document}